# Emergence of Language in the Developing Brain

**Linnea Evanson**[1,2,3], **Christine Bulteau**[2], **Mathilde Chipaux**[2], **Georg Dorfmüller**[2], **Sarah Ferrand-Sorbets**[2], **Emmanuel Raffo**[2], **Sarah Rosenberg**[2], **Pierre Bourdillon**[2,4*], **Jean-Rémi King**[1*]

[1]Meta AI, [2]Foundation Adolphe de Rothschild Hospital, Paris, France, [3]Laboratoire des systèmes perceptifs, Ecole Normale Supérieure, PSL University, CNRS, Paris, France, [4]Integrative Neuroscience & Cognition Center, Paris Cité University, Paris, France
*Equal contribution

A few million words suffice for children to acquire language. Yet, the brain mechanisms underlying this unique ability remain poorly understood. To address this issue, we investigate neural activity recorded from over 7,400 electrodes implanted in the brains of 46 children, teenagers, and adults for epilepsy monitoring, as they listened to an audiobook version of "The Little Prince". We then train neural encoding and decoding models using representations, derived either from linguistic theory or from large language models, to map the location, dynamics and development of the language hierarchy in the brain. We find that a broad range of linguistic features is robustly represented across the cortex, even in 2–5-year-olds. Crucially, these representations evolve with age: while fast phonetic features are already present in the superior temporal gyrus of the youngest individuals, slower word-level representations only emerge in the associative cortices of older individuals. Remarkably, this neuro-developmental trajectory is spontaneously captured by large language models: with training, these AI models learned representations that can only be identified in the adult human brain. Together, these findings reveal the maturation of language representations in the developing brain and show that modern AI systems provide a promising tool to model the neural bases of language acquisition.

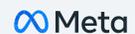

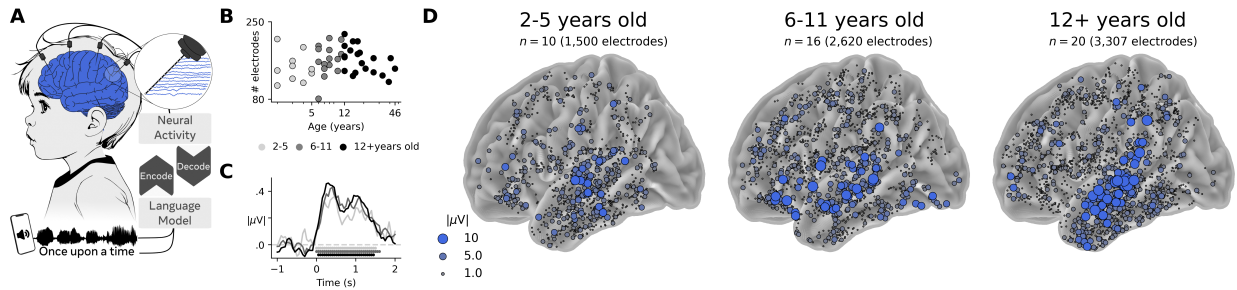

**Fig. 1| Approach to study the neural representations of language across development. A**. 46 participants implanted with 7,427 intracranial electrodes as part of their treatment for epilepsy, listened to an audiobook recording of *Le Petit Prince*, by Antoine de Saint Exupéry[1]. The neural representations of these brain signals are then analyzed in light of language models, either derived from linguistics or from modern AI systems. Encoding consists of linearly predicting brain responses to speech given language features, while decoding consists of the reverse objective. **B**. Number of electrodes for each participant. Colors indicate the three age groups used in subsequent analyses. **C**. Average broadband response, for each age group, time-locked to word onset. Horizontal lines indicate moments when the average response is above chance (p<0.05) across individuals. **D**. Average broadband response at 100 ms relative to word onset. For clarity, the left and right hemispheres are merged onto the left hemisphere, and electrodes are plotted to the closest cortical vertex (See Supplementary Fig.2 for additional visualization of the electrodes).

Human children have a remarkable capacity to acquire language. With just a few million words, they learn to categorize phonemes[2–4], combine them into a meaningful lexicon[5,6], and construct complex sentences[7,8]. In adults, the neural representations of language are increasingly understood: the auditory cortex generates brief, low-level features such as phonemes, while higher cortical regions, such as the Superior Temporal Gyrus (STG), Lateral / Inferior Frontal Cortex (L/IFC), and Inferior Parietal Lobule (IPL) represent word meanings and construct syntactic structures[9–15]. However, when and how the neural representations of language precisely emerge in the child brain over the course of development remains largely unknown[3,16–19].

This knowledge gap is caused by both technical and theoretical challenges. Technically, children's limited ability to remain still and engaged in repetitive tasks restrict classic functional Magnetic Resonance Imaging (fMRI) and Electroencephalography (EEG) paradigms[20–25]. Theoretically, models of language acquisition tend to focus on conceptual and behavioral predictions and often lack specificity regarding neural mechanisms and their development[3,26,27].

The recent advances in the field of Artificial Intelligence (AI) may help address this issue. Indeed, not only does machine learning facilitate the denoising and analysis of complex brain signals, but Large Language Models (LLMs) spontaneously learn representations that linearly map onto those of the adult brain[13,28–30]. Although these modern AI models are not designed for neuroscience, they provide an unprecedented tool to explore how language exposure may progressively shape the geometry of language representations in neural networks.

To test this possibility and precisely characterize the development of language representations in the brain, we use a variety of machine learning and AI algorithms to analyze and model a large set of pediatric intracranial recordings. Specifically, we recorded neural activity from 46 participants (aged 2 to 46 years) who had stereotactic EEG (sEEG) electrodes implanted for clinical purposes, as they listened to an audiobook of *Le Petit Prince*[1,31], leading to neural responses to 283 K words and 824 K phones recorded from over 7,400 electrodes distributed across the cortex (Fig.1A-B). Using encoding and decoding models targeting features either (i) predicted by theoretical linguistics or (ii) learned by LLMs, we map the emergence, timing, and localization of language representations across the cortex.

## Results

### Neural responses to speech in the child brain

*Average neural responses to speech.* To investigate the neural representations of natural speech across development, we first analyze the brain responses evoked at word and phoneme onset. To focus on spatially-resolved neural signals, we extract the Local Field Potential (LFP) of each bipolar construct of electrode pairs – hereafter labeled as "electrode" for simplicity. The results show that words (Fig.1C,D) and phonemes (Fig.S1) elicit responses in the brain areas typically associated with speech comprehension, namely, the STG, Superior Temporal Sulcus (STS), Anterior Temporal Lobe (ATL), IPL and L/IFC.

*Detecting linguistic representations with linear encoding.* To test whether these neural responses represent specific language features, we implemented a Temporal Response Function (TRF)[32,33] – a linear model optimized to predict the neural activity of each electrode from the classic linguistic features encompassing both phonetic and lexical features (Fig.2A). The phonetic features characterize the articulatory components of speech (*e.g.* voiced, plosive, nasal) irrespective of their meaning. By contrast, lexical features describe words, and here include the Zipf frequency[34] and part-of-speech categories (*e.g.* noun, verb, adjective). We evaluate this TRF encoding model with a held-out test set, by correlating its predicted neural responses with the true neural data. Overall, the results show that phonetic and lexical representations are best encoded in the speech cortex, and in particular in the STG (2-5 years: R=0.053, Wilcoxon test across participants: p=0.001 | 6-11 years: R=0.129, p<0.001 | 12+ years: R=0.116, p<0.001) and STS (2-5 years: R=0.040, p= 0.006 | 6-11 years: R=0.091, p<0.001 | 12+ years: R=0.08, p<0.001, Fig.2A-B). By contrast, areas outside of the language network, such as the occipital cortex, did not show significant encoding scores (2-5 years: R=0.008, p= 0.250 | 6-11 years: R=0.010, p=0.125 | 12+ years: R=0.011, p=0.078).

*Improvement of language representations across development.* With some notable exceptions (e.g. ATL and IPL), these neural representations of language can be significantly detected within each language Region of Interest (ROI) in each of the three age groups (Fig.2B). Critically, the TRF scores increase with development (Mann Whitney U test between the youngest group and the middle and older cohort led to p=0.039 and p=0.049, respectively). This effect, observed on average across the whole brain, appears to be most prominent in the STG, STS and L/IFC (Fig.2B). Overall, these results extend the findings reported in adults, by showing (i) that similar neural representations of language can be found in the brains of 2-5 year olds and (ii) that these neural representations evolve across childhood.

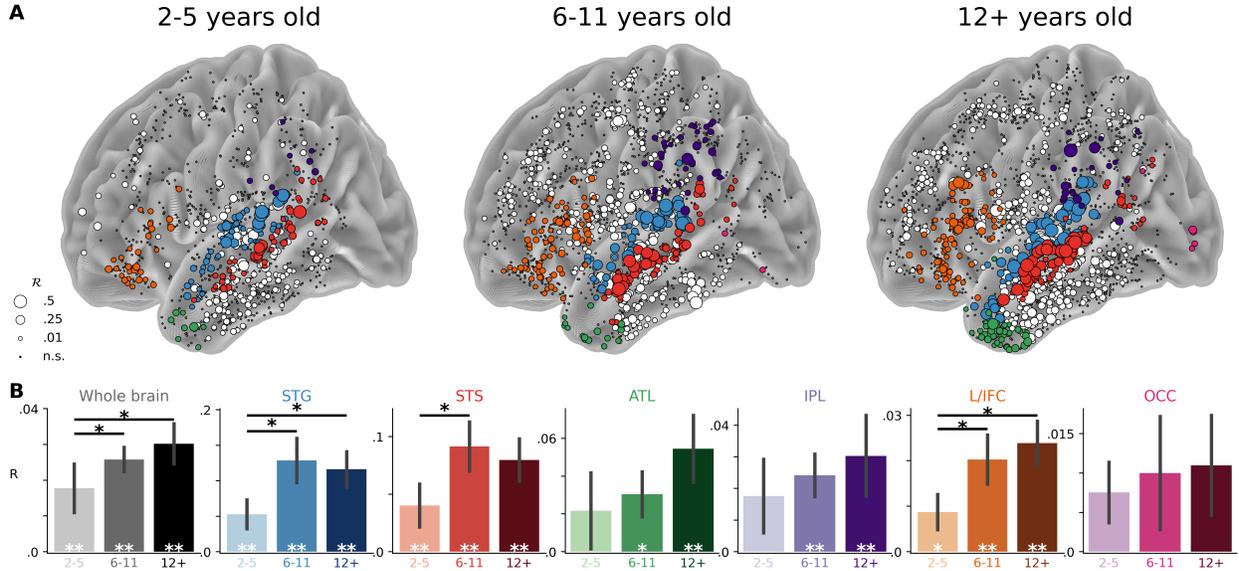

**Fig. 2| The neural representations of language change across development.** **A**. Encoding score for each electrode, as evaluated by linearly predicting neural activity from linguistic features ($R(phonemes+words)$). Black electrodes indicate that neural activity is not significantly predicted from linguistic features (n.s.). Colors indicate the regions of interest, as identified within each subject. **B**. Average encoding score for each of the three age groups, for the whole brain, and for each region of interest. Error bars indicate the Standard Error of the Mean (SEM) across subjects. White and black stars indicate statistical significance within and across age groups, respectively.

## Hierarchical organization of language representations across development

*Anatomical organization of language representations.* To clarify the anatomical organization of these language representations, we implemented a second TRF model trained, this time, with phonetic features only. We then compare its encoding scores ($R(phonemes)$) to those of the TRF model described above ($R(phonemes + words)$). The resulting comparison isolates the gain in neural predictions brought by lexical features: $\Delta R(words) = R(phonemes + words) - R(phonemes)$. The results show that phonetic representations peak in STG, but are significant across a large set of brain regions (Fig.3A). In comparison, word representations appear to be more homogeneously distributed across the cortex (Fig.3D), with particular improvement in ATL (p=0.0290, Fig.S3).

*Time course of language representations.* To clarify the dynamics of these phonetic and lexical representations, we next implemented a time-resolved decoding model. Specifically, we train, within each ROI of each participant, a linear model to predict phonetic or lexical features from the neural activity relative to phoneme or word onsets, respectively (Fig.3B,E). The results show that phonetic features rise slightly before phoneme onset, peak around 150 ms and rapidly drop back to chance level (Fig.3B). By contrast, lexical features begin rising up to 1 s before word onset, peak around 350 ms, and show a sustained response lasting up to 2 s after word onset (Fig.3E).

*Development of language representations.* To evaluate how these representations evolve during development, we separately analyzed decoding results within each age group. Overall, a broad set of language features are represented in neural activity in each age group (Fig.3C,F). However, and unlike the two older groups, only a subset of phonetic and lexical features are detected in 2-5 year olds (Fig.3C, F). Statistical comparisons across age groups further show that while phonetic representations do not significantly vary across age (Mann Whitney U test between the cohorts 2-5:6-11 p=0.128, 2-5:12+ p=0.387, and 6-11:12+ p=0.643), lexical representations significantly improve across development (2-5:6-11 p=0.0387, 2-5:12+ p=0.0339 and 6-11:12+ p=0.569, Fig.3C,F, Fig.4A,D).

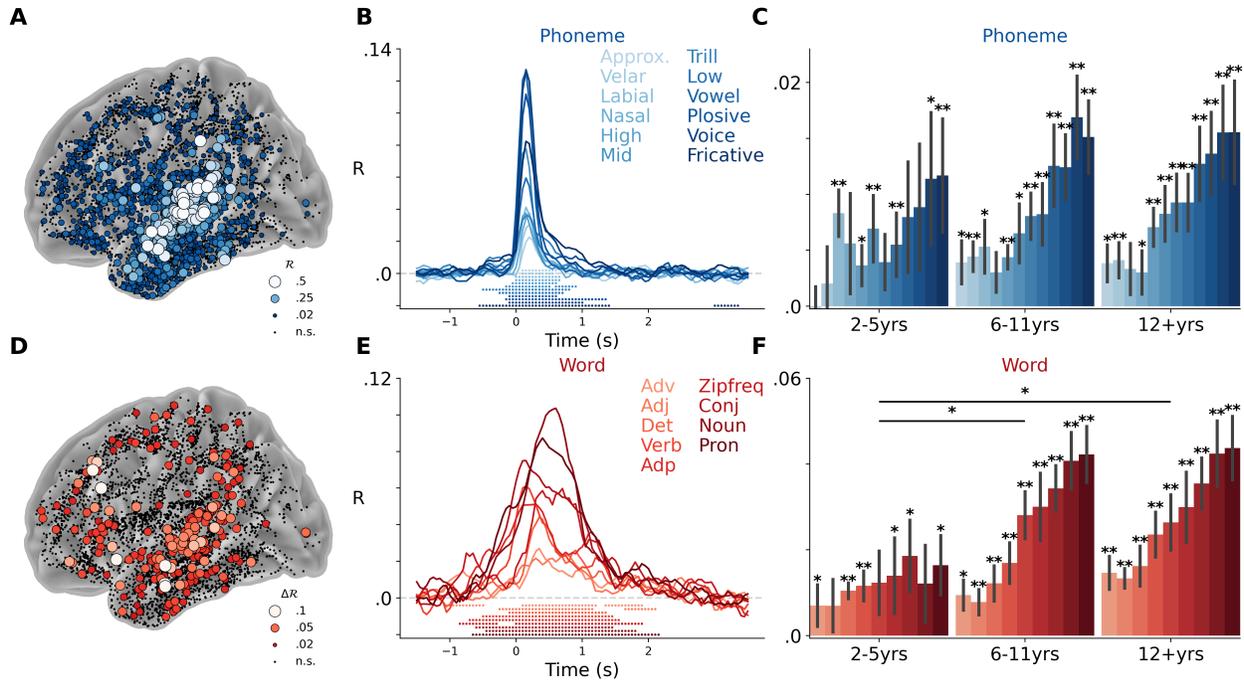

**Fig. 3| Anatomical, temporal and developmental bases of the language hierarchy** **A**. Encoding scores of a model fit on phonetic features only, on each responsive electrode. **B**. Whole-brain decoding scores of each phonetic feature relative to phoneme onset, averaged over all participants. Horizontal lines indicate significant time-steps. **C**. Average decoding score between [-1 s, 2 s] for each phonetic feature. Error bars indicate SEM across subjects. Stars indicate significance across subjects. **D**. Gain of encoding scores obtained by adding word-features to the encoding model used in A: $\Delta R = R(word + phoneme) - R(phoneme)$. **E-F**. As in B-C, for lexical features.

*Development of the language network.* To further clarify (i) the presence and (ii) the evolution of phonetic and lexical representations within each brain region, we trained a distinct linear decoder for each ROI of each participant. First, the dynamics of phonetic and lexical representations were similar across the three age groups, except that word-level representations tended to be attenuated in the youngest individuals as compared to older ones (Fig.4). Notably, this led the younger groups lacking detectable representations of language in IPL, L/IFC and ATL, in spite of demonstrating clear language representations in STG, with significant differences compared to older age groups in the L/IFC (word level: 2-5:12+ p=0.0237) and ATL (word level: 2-5:12+ p=0.0440) in particular. These results suggest that, as the brain matures, language features are broadcast to and represented in an increasingly large cortical network.

## AI models account for the developmental changes of language representations

*Using AI to model the geometry of language representations.* The phonetic and lexical representations explored above derive from the descriptive features of linguistic models. Can *trainable* language models further account, not only for the presence and structure of language representations in the brain, but also their developmental trajectory? Specifically, does training an LLM lead it to learn representations that are only observable in adults? To explore this hypothesis, we apply our encoding and decoding analyses using, this time, the representations of two landmark architectures, before and after their training. The first model, wav2vec 2.0[35] is trained on 53 k hours of speech *sounds*[1]. The second, Llama 3.1[39] is a large transformer trained on 15.6 trillion *textual* tokens. To facilitate computing and comparison, we focus, for each of these models, on two representative layers and transform their high-dimensional ac-tivations into 50 principal components (Fig.5A,E).

---
[1]This represents an equivalent of approximately 500 M textual tokens, given an estimate of 150 tokens per minute.

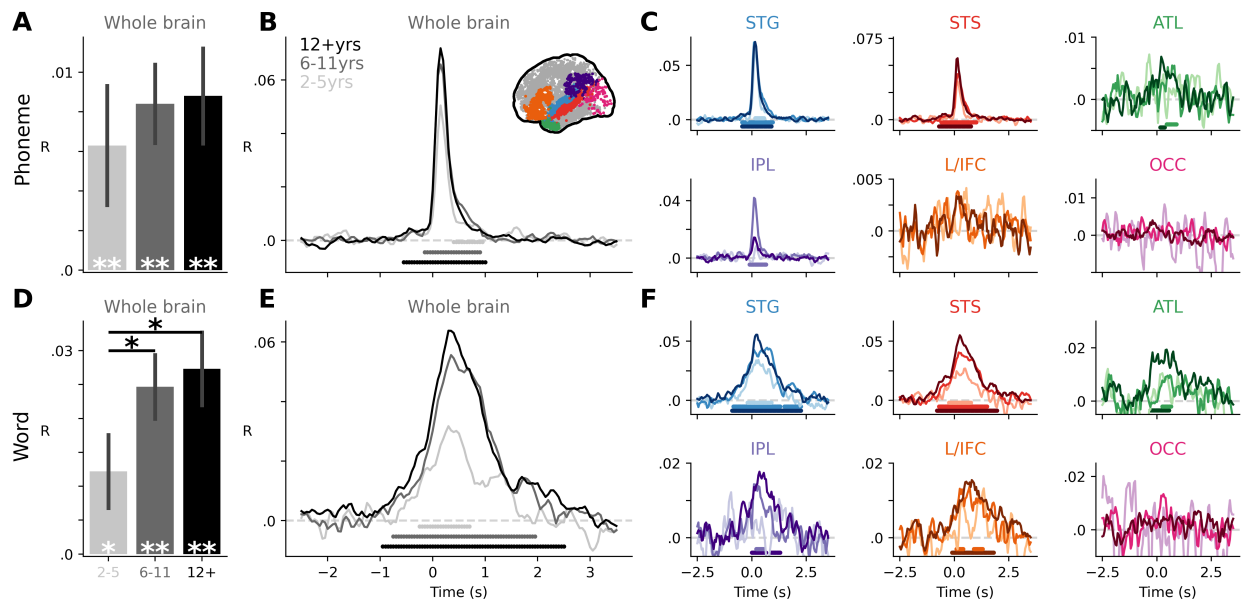

**Fig. 4| Evolution of word-level representations across childhood.** **A**. Average phonetic decoding per age group between [-1s, 2s] relative to phoneme onset. Error bars are SEM across participants. White and black stars indicate within- and across-groups statistical significance, respectively. **B**. Phonetic decoding time courses relative to phoneme onset. Horizontal lines indicate statistically significant time-steps. **C**. Same as B, for each region of interest. **D-F**. Same as A-C for lexical decoding relative to word onset.

*Training makes speech and large language models more similar to the brain.* First, and as previously reported in adults[29,36,37], *untrained* models already account for neural responses in the speech cortex (Fig.5B,C,F,G). Second, training these AI models leads to a gain in encoding scores for both wav2vec 2.0 and Llama 3.1 (Fig.5B,F). This gain in brain-predictability can be observed within each age group, and the largest effects peak in brain areas associated with speech processing: STG and STS (Fig.5C,G). Together, these results show that training these language models make their activations more similar to those of the brain.

*AI models capture the hierarchical organization of language representations in the cortex.* Can LLMs also account for the anatomical hierarchy of language? To test this question, we first compare the encoding scores of deep ($L_{0.8}$) and shallow layers ($L_{0.1}$). For wav2vec 2.0, this comparison leads to a wide-spread gain in encoding performance (mean $\Delta_L R=0.0113$, p<10e8 across participants), peaking in the brain areas associated with language processing, and in particular in STG and STS, ATL and IPL (Fig.5C). For Llama 3.1, however, we do not observe such gain (mean $\Delta_L R=-0.001$, p=0.929). Second, to clarify the neural dynamics associated with these AI representations, we implement linear decoding models. Similarly to phonemes *versus* words (Fig.3B,E), the results show that the deep layer of wav2vec 2.0 is significantly aligned with brain activity for a longer time period than its shallow layer (Fig.5D). However, this difference in neural dynamics is, again, not observed for the representations of Llama 3.1 (Fig.5H). Similarly to lexical features (Fig.3E), the neural time courses associated with the shallow and deep layers of Llama 3.1 were both long lasting (Fig.5H). We speculate that this lack of hierarchical effects may be caused by the fact that the first layer of Llama 3.1 is already representing word-level information.

*A shared developmental trajectory between LLMs and the brain?* Finally, to test whether the training trajectories of these language models qualitatively account for the evolution of neural representations across development, we compare decoding across age groups using the representations of the best-performing layer of each model. The neural decoding of wav2vec 2.0 representations did not clearly vary with age (2-5:6-11 p=0.060, 2-5:12+ p=0.234, 6-11:12+ p=0.722, Fig.6A–D). although an age-dependent effect is observable when compar-ing the hierarchy of representations learned by the model, $\Delta_L R$ (2-5:6-11 p=0.0307, 2-5:12+ p=0.0495, 6-11:12+ p=0.700, Fig.5B and Fig.S4). In contrast, the neural decoding of Llama 3.1 representations significantly improved with age (2-5:6-11 p=0.0187, 2-5:12+ p=0.0308, 6-11:12+ p=0.569, Fig.6 F-I). To further test whether this gain in decoding performance is caused by acquired representations, we evaluate how the training gain ($\Delta_T R = R_{Trained} - R_{Random}$) varies across age group. As before, we observe that Llama 3.1 specifically learns representations that can only be observed in the older groups ($\Delta_T R_{12+} > \Delta_T R_{2-5}$, pval=0.0411 and $\Delta_T R_{6-11} > \Delta_T R_{2-5}$, pval=0.0212 Fig.6E,J). Together, these results suggest that the changes in language representations induced by training this LLM partially capture the changes in language representations in the developing brain.

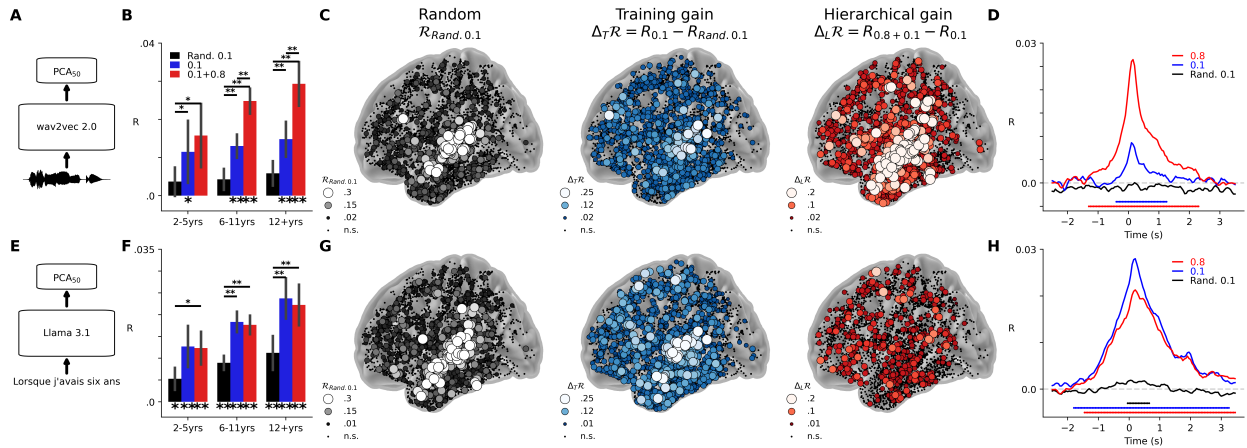

**Fig. 5| Speech and large language models account for the dynamics and hierarchy of neural representations.** **A**. Wav2vec 2.0[35] is a self-supervised model trained on trained 53k hours of speech sounds, and known to learn representations aligned to the adult brain[36–38]. We investigate both trained and randomly initialized embeddings, extracted from two layers of wav2vec 2.0, located at 10% and 80% of its hierarchy, respectively. All embeddings are then PCA transformed into 50 orthogonal components. **B**. Average encoding scores (across all electrodes) for each age group, and each wav2vec 2.0 embedding. Error bars indicate SEM across subjects. Below and top stars indicate within- and across-group significance, respectively. **C**. Encoding scores ($R$) and encoding gains ($\Delta R$) for each electrode as estimated with (the comparison of) TRF model(s) input with one of the embeddings of wav2vec 2.0. **D**. Decoding time course, relative to word onset, for each of the three wav2vec 2.0 embeddings. Horizontal lines indicate statistically significant time-steps. **E-H**. Same as A-D, for Llama 3.1[39], an LLM trained on 15.6 trillion textual tokens.

## Discussion

*Results summary.* This study provides a detailed characterization of the evolution of language representations in the brain, from the cortical responses of young children, teenagers and adults, as they listened to natural speech. Our encoding and decoding results show that a rich set of language features is already present in the cortex of 2–5-year-olds, with neural dynamics comparable to adults, but initially restricted to a narrow set of regions that gradually expand during development. These anatomico-functional changes are accompanied with a development of the hierarchy of language representations: while low-level phonetic features are consistently represented from 2 years old to adulthood, several high-level lexical features only emerge in the cortical activity of individuals aged over 6 years old. Finally, these findings, derived from the representations of linguistic theory, are partially captured by the representations learned by wav2vec 2.0[35] and Llama 3.1[39], two landmark AI models, respectively trained on speech and text. For clarity, we first discuss the results derived from linguistic models, before turning to those derived from AI models.

*How linguistic representations develop in the child brain.* The temporal and anatomical profiles of language representations here identified in the brains of young children are remarkably consistent with those of the adult brain[11,14,40–42]. Together, these results strengthen and extend the view that core compo-nents of language processing are already functional in the early phase of development[21,43–47]. Furthermore, high-level processing stages of language appear to slowly mature throughout childhood especially in the high-level cortices (e.g., STS, IPL). This finding aligns with both (i) the slow anatomical maturation of associative cortices[48] and (ii) hierarchical theories of language acquisition, in which low-level representations, such as phonemes, are acquired during infancy, while higher-level features such as words and semantics emerge more gradually[2,4,26,49]. Importantly, our approach complements these findings by providing, through the direct tracking of representational changes across development, new empirical elements to inform and constrain the theories of language acquisition.

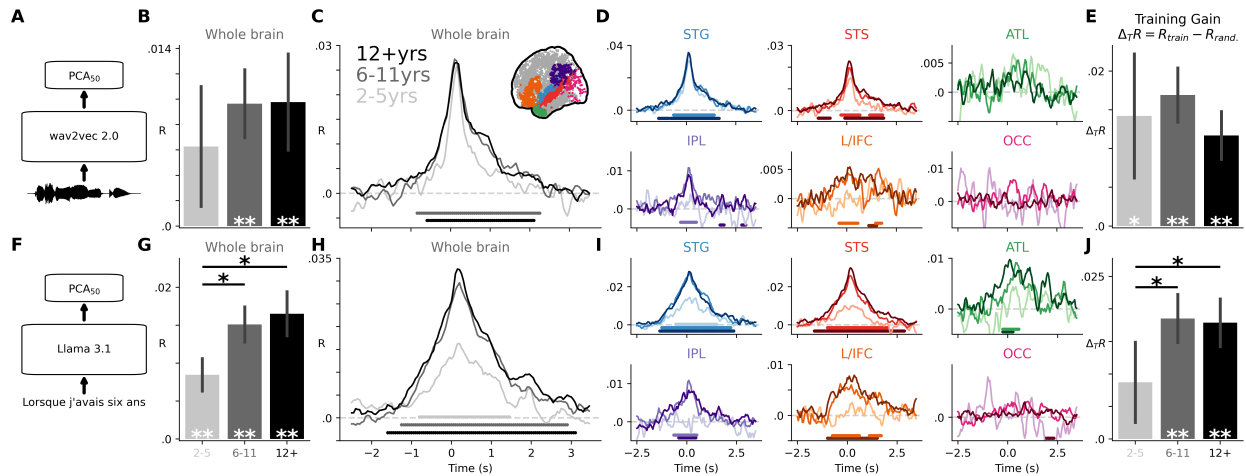

**Fig. 6| Training speech and large language models account for the developmental trajectory of language representations in the brain. A**. Wav2vec 2.0[35], same as Fig.5. Results shown for the layer best decoded by this model in Fig.5. **B**. Average decoding scores across the whole brain between [-1s, 2s] relative to word onset. Error bars indicate SEM across subjects. White and black stars indicate significant within- and across-groups, respectively. **C**. Decoding scores across time, for each age group. Dots indicate statistically significant time-steps. **D**. Same as C, for each region of interest separately. **E**. Training gain $\Delta_T R$, averaged over all electrodes in the language ROIs (STG, STS, ATL, IPL, L/IFC). Error bars indicate SEM across subjects. **F-J**. Same as A-E, with Llama 3.1[39].

*Modeling the emergence of language representations in the brain with AI.* If language representations in the brain emerge hierarchically, a key question remains: can this emergence be captured by computational models? Here, we show that modern AI models offer surprisingly useful tools to address this issue. Unlike descriptive and conceptual models, these algorithms make precise predictions about (i) the geometry of language representations and (ii) their evolution with language exposure. In addition, and while these algorithms differ substantially from biological neural networks, they generate representations that linearly align with those of the adult brain recorded with Electrocorticography (ECoG)[38,50], fMRI[30,36,51–54], and Magnetoencephalography (MEG)[53]. The present study complements these adult studies by showing similarities between LLMs and the developing human brain, in four main ways. First, both wav2vec 2.0[35] and Llama 3.1[39] learn representations that align with those of children, teenagers and adults (Fig.5B,C,F,G). Second, wav2vec 2.0 develops a layer-wise structure that mirrors the anatomical and temporal hierarchy of language processing in the brain. This property, however, is not observed in Llama 3.1, in which the first layer already encodes word-level information. Third, training these models improves their representational similarity with the developing brain (Fig.5B,F). Forth, the representations acquired by Llama 3.1 through training can only be observed in the brain of the oldest individuals (Fig.6G-J). Together, these results indicate that language representations evolve in similar ways in the developing human brain and in modern language models during training. An important next step will now consist of elucidating the key factors – model architectures and training regimes – and the type of neural codes[55–58] that drive this partial convergence between modern AI systems and the devel-

oping brain[27,59,60].

*Limitations.* The present work should be interpreted in light of several important limitations. First, our cohort consists of French-speaking children, who suffer from severe epilepsy. Consequently, the present results may not be representative of the evolution of language in the healthy brain[61–63]. Second, although our sample size is relatively large (n=46) compared to intracranial studies exploring the high-dimensional representations of speech in adults (e.g., Mesgarani & Chang[64]: n=3; Mesgarani *et al.*[11]: n=6; Flinker *et al.*[65]: n=8; Hamilton *et al.*[66]: n=9; Thomas *et al.*[42]: n=8; Li *et al.*[38]: n=12; Giroud *et al.*[15]: n=11; Goldstein *et al.*[50]: n=9), it remains modest for investigating inter-individual differences across development[67]. In particular, this sample size limits statistical power, and with it, our ability to clarify subtle developmental stages – such as those elicited by literacy around age six[19]. Third, the first months of life are known to be critical for speech development[2,4,27]. However, these early developmental stages remain inaccessible to intracranial studies, as sEEG is rarely performed in children under two years of age due to the insufficient rigidity of their skull[68]. Fourth, while audiobooks are naturalistic and engaging, they limit our ability to disentangle rare but theoretically important linguistic phenomena. In particular, isolating structures such as long-distance dependencies, syntactic constituents and reference[69,70], or extending this effort to language production[71] remain important future challenges.

*Conclusion* Finally, LLMs remain highly imperfect models: their architectures benefit from an implausibly large and precise working memory, they still struggle with various linguistic tasks[72–74] and require a considerably larger exposure to language than humans[59,60,73,75]. This inefficiency underscores the fundamental differences in how artificial and human neural systems acquire knowledge. Despite these clear differences, however, the present findings point to computational principles unexpectedly shared between biological and artificial systems, offering a potential blueprint for building a unified model of language processing and acquisition in the human brain[53,76,77].

1. De Saint-Exupéry, A. *Le Petit Prince* French. First edition (Reynal & Hitchcock, New York, 1943).
2. Kuhl, P. K., Williams, K. A., Lacerda, F., Stevens, K. N. & Lindblom, B. Linguistic experience alters phonetic perception in infants by 6 months of age. *Science* **255,** 606–608 (1992).
3. Kuhl, P. K. Early language acquisition: cracking the speech code. *Nature reviews neuroscience* **5,** 831–843 (2004).
4. Werker, J. F. & Hensch, T. K. Critical periods in speech perception: new directions. *Annual review of psychology* **66,** 173–196 (2015).
5. Jusczyk, P. W. & Hohne, E. A. Infants' memory for spoken words. *Science* **277,** 1984–1986 (1997).
6. Bergelson, E. & Swingley, D. At 6–9 months, human infants know the meanings of many common nouns. *Proceedings of the National Academy of Sciences* **109,** 3253–3258 (2012).
7. Kovács, A. M. & Endress, A. D. Hierarchical processing in seven-month-old infants. *Infancy* **19,** 409–425 (2014).
8. Romeo, R. R. *et al.* Beyond the 30-million-word gap: Children's conversational exposure is associated with language-related brain function. *Psychological science* **29,** 700–710 (2018).
9. Price, C. J. The Anatomy of Language: A Review of 100 fMRI Studies Published in 2009. *Annals of the New York Academy of Sciences* **1191,** 62–88. ISSN: 1749-6632 (2010).
10. Pallier, C., Devauchelle, A.-D. & Dehaene, S. Cortical representation of the constituent structure of sentences. *Proceedings of the National Academy of Sciences* **108,** 2522–2527 (2011).
11. Mesgarani, N., Cheung, C., Johnson, K. & Chang, E. F. Phonetic feature encoding in human superior temporal gyrus. *Science* **343,** 1006–1010 (2014).
12. Caucheteux, C., Gramfort, A. & King, J.-R. Disentangling syntax and semantics in the brain with deep networks. *ICML,* 1336–1348 (2021).
13. Huth, A. G., De Heer, W. A., Griffiths, T. L., Theunissen, F. E. & Gallant, J. L. Natural speech reveals the semantic maps that tile human cerebral cortex. *Nature* **532,** 453–458 (2016).
14. Keshishian, M. *et al.* Joint, Distributed and Hierarchically Organized Encoding of Linguistic Features in the Human Auditory Cortex. *Nature human behaviour* **7,** 740–753. ISSN: 2397-3374 (2023).
15. Giroud, J., Trébuchon, A., Mercier, M., Davis, M. H. & Morillon, B. The human auditory cortex concurrently tracks syllabic and phonemic timescales via acoustic spectral flux. *Science Advances* **10** (2024).
16. Cheour, M. *et al.* Development of language-specific phoneme representations in the infant brain. *Nature neuroscience* **1,** 351–353 (1998).
17. Zatorre, R. J. Predispositions and plasticity in music and speech learning: neural correlates and implications. *Science* **342,** 585–589 (2013).
18. Skeide, M. A. & Friederici, A. D. The ontogeny of the cortical language network. *Nature reviews neuroscience* **17,** 323–332 (2016).
19. Dehaene, S. *How we learn: The new science of education and the brain* (Penguin UK, 2020).
20. Dehaene-Lambertz, G., Dehaene, S. & Hertz-Pannier, L. Functional Neuroimaging of Speech Perception in Infants. *Science* **298,** 2013–2015. ISSN: 00368075 (2002).
21. Dehaene-Lambertz, G. *et al.* Functional organization of perisylvian activation during presentation of sentences in preverbal infants. *Proceedings of the National Academy of Sciences* **103,** 14240–14245 (2006).

22. Shultz, S., Vouloumanos, A., Bennett, R. H. & Pelphrey, K. Neural specialization for speech in the first months of life. *Developmental science* **17,** 766–774 (2014).
23. Di Liberto, G. M. *et al.* Emergence of the cortical encoding of phonetic features in the first year of life. *Nature communications* **14,** 7789 (2023).
24. Mersad, K., Kabdebon, C. & Dehaene-Lambertz, G. Explicit access to phonetic representations in 3-month-old infants. *Cognition* **213,** 104613 (2021).
25. Gennari, G., Marti, S., Palu, M., Fló, A. & Dehaene-Lambertz, G. Orthogonal Neural Codes for Speech in the Infant Brain. *Proceedings of the National Academy of Sciences* **118** (2021).
26. Friedmann, N., Belletti, A. & Rizzi, L. Growing trees: The acquisition of the left periphery. *Glossa: a journal of general linguistics* **6** (2021).
27. Dupoux, E. Cognitive Science in the Era of Artificial Intelligence: A Roadmap for Reverse-Engineering the Infant Language-Learner. *Cognition* **173,** 43–59. ISSN: 00100277 (2018).
28. Mitchell, T. M. *et al.* Predicting Human Brain Activity Associated with the Meanings of Nouns. *Science* **320,** 1191–1195 (2008).
29. Schrimpf, M. *et al.* The neural architecture of language: Integrative modeling converges on predictive processing. *Proceedings of the National Academy of Sciences* **118,** e2105646118 (2021).
30. Caucheteux, C., Gramfort, A. & King, J.-R. Evidence of a predictive coding hierarchy in the human brain listening to speech. *Nature human behaviour* **7,** 430–441 (2023).
31. Li, J. *et al.* Le Petit Prince Multilingual Naturalistic fMRI Corpus. *Scientific Data* **9,** 530. ISSN: 2052-4463 (2022).
32. Theunissen, F. *et al.* Estimating Spatio-Temporal Receptive Fields of Auditory and Visual Neurons from Their Responses to Natural Stimuli. *Network: Computation in Neural Systems* **12,** 289–316. ISSN: 0954-898X, 1361-6536 (2001).
33. Lalor, E. C. & Foxe, J. J. Neural responses to uninterrupted natural speech can be extracted with precise temporal resolution. *European journal of neuroscience* **31,** 189–193 (2010).
34. Zipf, G. K. Human behavior and the principle of least effort: An introduction to human eoclogy (1949).
35. Baevski, A., Zhou, H., Mohamed, A. & Auli, M. Wav2vec 2.0: A Framework for Self-Supervised Learning of Speech Representations. *Advances in Neural Information Processing Systems* **2020-Decem,** 1–12. eprint: 2006.11477 (2020).
36. Millet, J. *et al.* Toward a realistic model of speech processing in the brain with self-supervised learning. *Advances in Neural Information Processing Systems* **35,** 33428–33443 (2022).
37. Vaidya, A. R., Jain, S. & Huth, A. *Self-Supervised Models of Audio Effectively Explain Human Cortical Responses to Speech* in *Proceedings of the 39th International Conference on Machine Learning* (PMLR, 2022), 21927–21944.
38. Li, Y. *et al.* Dissecting neural computations in the human auditory pathway using deep neural networks for speech. *Nature Neuroscience* **26,** 2213–2225 (2023).
39. Grattafiori, A. *et al. The Llama 3 Herd of Models* 2024. eprint: 2407.21783 (cs). (2025).
40. De Heer, W. A., Huth, A. G., Griffiths, T. L., Gallant, J. L. & Theunissen, F. E. The hierarchical cortical organization of human speech processing. *Journal of Neuroscience* **37,** 6539–6557 (2017).
41. Forseth, K. J. *et al.* A lexical semantic hub for heteromodal naming in middle fusiform gyrus. *Brain* **141,** 2112–2126 (2018).
42. Thomas, T. M. *et al.* Decoding articulatory and phonetic components of naturalistic continuous speech from the distributed language network. *Journal of neural engineering* **20,** 046030 (2023).
43. Ahmad, Z., Balsamo, L., Sachs, B., Xu, B. & Gaillard, W. Auditory comprehension of language in young children: neural networks identified with fMRI. *Neurology* **60,** 1598–1605 (2003).
44. Brusini, P., Dehaene-Lambertz, G., Dutat, M., Goffinet, F. & Christophe, A. ERP evidence for on-line syntactic computations in 2-year-olds. *Developmental Cognitive Neuroscience* **19,** 164–173 (2016).
45. Arya, R. *et al.* Development of information sharing in language neocortex in childhood-onset drug-resistant epilepsy. *Epilepsia* **60,** 393–405 (2019).
46. Ozernov-Palchik, O. *et al.* Precision fMRI Reveals That the Language Network Exhibits Adult-like Left-Hemispheric Lateralization by 4 Years of Age. *bioRxiv,* 2024.05.15.594172. ISSN: 2692-8205 (2024).
47. Kanno, A. *et al.* Dynamic Causal Tractography Analysis of Auditory Descriptive Naming: An Intracranial Study of 106 Patients. *bioRxiv,* 2025.03.07.641428 (2025).
48. Gogtay, N. *et al.* Dynamic mapping of human cortical development during childhood through early adulthood. *Proceedings of the national academy of sciences* **101,** 8174–8179 (2004).
49. Schipke, C. S., Knoll, L. J., Friederici, A. D. & Oberecker, R. Preschool children's interpretation of object-initial sentences: Neural correlates of their behavioral performance. *Developmental Science* **15,** 762–774 (2012).
50. Goldstein, A. *et al.* Shared computational principles for language processing in humans and deep language models. *Nature neuroscience* **25,** 369–380 (2022).
51. Jain, S. & Huth, A. Incorporating context into language encoding models for fMRI. *Advances in neural information processing systems* **31** (2018).
52. Toneva, M. & Wehbe, L. Interpreting and improving natural-language processing (in machines) with natural language-processing (in the brain). *Advances in neural information processing systems* **32** (2019).
53. Caucheteux, C. & King, J.-R. Brains and algorithms partially converge in natural language processing. *Communications biology* **5,** 134 (2022).
54. Pasquiou, A., Lakretz, Y., Thirion, B. & Pallier, C. Information-restricted neural language models reveal different brain regions' sensitivity to semantics, syntax, and context. *Neurobiology of Language* **4,** 611–636 (2023).
55. Hewitt, J. & Manning, C. D. A structural probe for finding syntax in word representations. *Proceedings of the 2019 Conference of the North American Chapter of the Association for Computational Linguistics: Human Language Technologies, Volume 1 (Long and Short Papers),* 4129–4138 (2019).
56. Lakretz, Y. *et al.* The Emergence of Number and Syntax Units in LSTM Language Models. *Proceedings of the 2019 Conference of the North American Chapter of the Association for Computational Linguistics: Human Language Technologies, Volume 1 (Long and Short Papers),* 11–20 (2019).
57. Diego Simon, P. J., d'Ascoli, S., Chemla, E., Lakretz, Y. & King, J.-R. A polar coordinate system represents syntax in large language models. *Advances in Neural Information Processing Systems* **37,** 105375–105396 (2024).

58. Evanson, L., Lakretz, Y. & King, J. R. Language Acquisition: Do Children and Language Models Follow Similar Learning Stages? *Findings of the Association for Computational Linguistics: ACL 2023,* 12205–12218 (2023).
59. Warstadt, A. *et al.* Findings of the BabyLM Challenge: Sample-Efficient Pretraining on Developmentally Plausible Corpora. *Proceedings of the BabyLM Challenge at the 27th Conference on Computational Natural Language Learning,* 1–34 (2023).
60. Frank, M. C. & Goodman, N. D. Cognitive modeling using artificial intelligence (2025).
61. Haseeb, A. *et al.* Young patients with focal seizures may have the primary motor area for the hand in the postcentral gyrus. *Epilepsy research* **76,** 131–139 (2007).
62. Holmes, G. L. Effect of seizures on the developing brain and cognition. *Seminars in pediatric neurology* **23,** 120–126 (2016).
63. Flinker, A., Piai, V. & Knight, R. T. Intracranial electrophysiology in language research (2018).
64. Mesgarani, N. & Chang, E. F. Selective cortical representation of attended speaker in multi-talker speech perception. *Nature* **485,** 233–236 (2012).
65. Flinker, A., Doyle, W. K., Mehta, A. D., Devinsky, O. & Poeppel, D. Spectrotemporal modulation provides a unifying framework for auditory cortical asymmetries. *Nature human behaviour* **3,** 393–405 (2019).
66. Hamilton, L. S., Oganian, Y., Hall, J. & Chang, E. F. Parallel and distributed encoding of speech across human auditory cortex. *Cell* **184,** 4626–4639 (2021).
67. Hervé, E., Mento, G., Desnous, B. & François, C. Challenges and new perspectives of developmental cognitive EEG studies. *NeuroImage* **260,** 119508 (2022).
68. Katz, J., Armstrong, C., Kvint, S. & Kennedy, B. C. Stereoelectroencephalography in the very young: Case report. *Epilepsy & Behavior Reports* **19,** 100552 (2022).
69. Ding, N., Melloni, L., Zhang, H., Tian, X. & Poeppel, D. Cortical tracking of hierarchical linguistic structures in connected speech. *Nature neuroscience* **19,** 158–164 (2016).
70. Dijksterhuis, D. E. *et al.* Pronouns reactivate conceptual representations in human hippocampal neurons. *Science* **385,** 1478–1484 (2024).
71. Tourville, J. A. & Guenther, F. H. The DIVA model: A neural theory of speech acquisition and production. *Language and cognitive processes* **26,** 952–981 (2011).
72. Lakretz, Y., Desbordes, T., Hupkes, D. & Dehaene, S. Can transformers process recursive nested constructions, like humans? *Proceedings of the 29th International Conference on Computational Linguistics,* 3226–3232 (2022).
73. Srivastava, A. *et al.* Beyond the imitation game: Quantifying and extrapolating the capabilities of language models. *arXiv* (2022).
74. Hu, J. & Levy, R. Prompting Is Not a Substitute for Probability Measurements in Large Language Models. *Proceedings of the 2023 Conference on Empirical Methods in Natural Language Processing,* 5040–5060 (2023).
75. Gilkerson, J. *et al.* Mapping the early language environment using all-day recordings and automated analysis. *American journal of speech-language pathology* **26,** 248–265 (2017).
76. Huh, M., Cheung, B., Wang, T. & Isola, P. Position: The Platonic Representation Hypothesis. *Proceedings of the 41st International Conference on Machine Learning. Proceedings of Machine Learning Research* **235,** 20617–20642 (2024).
77. Hosseini, E. *et al.* Universality of representation in biological and artificial neural networks. *bioRxiv* (2024).
78. Hoff, E. & Shatz, M. *Blackwell handbook of language development* (John Wiley & Sons, 2009).
79. Fischl, B. FreeSurfer. *NeuroImage. 20 YEARS OF fMRI* **62,** 774–781. ISSN: 1053-8119 (2012).
80. Reuter, M., Schmansky, N. J., Rosas, H. D. & Fischl, B. Within-Subject Template Estimation for Unbiased Longitudinal Image Analysis. *NeuroImage* **61,** 1402–1418. ISSN: 1095-9572 (2012).
81. *CURRY Neuro Imaging Suite for Epilepsy Evaluation* https://compumedicsneuroscan.com/curry-epilepsy-evaluation/. Accessed: 21/02/2024.
82. Rockhill, A. P. *et al.* Intracranial Electrode Location and Analysis in MNE-Python. *Journal of Open Source Software* **7,** 3897. ISSN: 2475-9066 (2022).
83. Fedorenko, E., Ivanova, A. A. & Regev, T. I. The Language Network as a Natural Kind within the Broader Landscape of the Human Brain. *Nature Reviews Neuroscience,* 1–24. ISSN: 1471-0048 (2024).
84. Desikan, R. S. *et al.* An Automated Labeling System for Subdividing the Human Cerebral Cortex on MRI Scans into Gyral Based Regions of Interest. *NeuroImage* **31,** 968–980. ISSN: 1053-8119 (2006).
85. Destrieux, C., Fischl, B., Dale, A. & Halgren, E. Automatic Parcellation of Human Cortical Gyri and Sulci Using Standard Anatomical Nomenclature. *NeuroImage* **53,** 1–15. ISSN: 1053-8119 (2010).
86. Gramfort, A. *et al.* MEG and EEG Data Analysis with MNE-Python. *Frontiers in Neuroscience* **7,** 1–13 (2013).
87. Speer, R. *rspeer/wordfreq: v3.0* version v3.0.2. Accessed: 12/05/2025. https://doi.org/10.5281/zenodo.7199437.
88. *Space: Available trained pipelines for French* https://spacy.io/models/fr. Accessed: 21/02/2024.
89. *Scikit-Learn: Machine Learning in Python — Scikit-Learn 1.6.1 Documentation* Accessed: 09/05/2025. https://scikit-learn.org/stable/index.html.
90. *Wav2Vec2-Base Model Card* https://huggingface.co/facebook/wav2vec2-base. Accessed: 21/02/2024. 2024.
91. *Meta-Llama/Llama-3.1-8B* https://huggingface.co/meta-llama/Llama-3.1-8B. Accessed: 2025-05-09.
92. Benjamini, Y. & Hochberg, Y. Controlling the False Discovery Rate: A Practical and Powerful Approach to Multiple Testing. *Journal of the Royal Statistical Society: Series B (Methodological)* **57,** 289–300. ISSN: 2517-6161 (1995).

*Acknowledgments* We would like to thank all the participants and their families for their participation in this study. We thank the doctors, technicians and nurses at the Rothschild Foundation Hospital for their support of this project and for their comments and suggestions. This project has received funding from the European Union's Horizon 2020 research and innovation program under the Marie Skłodowska-Curie grant agreement No 945304, for L.E for her work at PSL. This work was granted access to the HPC resources of IDRIS under the allocation 2024-AD011013683R2 made by GENCI.

*Author contributions* L.E. implemented the experiment, collected, analysed data, wrote and reviewed the manuscript. C.B, M.C, G.D, S.F.S, E.R., S.R. collected data and reviewed the manuscript. P.B. conceived the study, collected data and reviewed the manuscript. J-R.K. conceived the study, directed the line of research, wrote and reviewed the manuscript.

*Competing interests* No authors declare any competing interests.

*Data and materials availability* The data that support the findings of this study are not openly available because of sensitivity but are available upon reasonable request. The analysis and data visualization code are available upon request.

*Supplementary Materials*

Materials and methods

Figs. S1 - S4

*Acronyms*

**AI**   Artificial Intelligence
**LLM**   Large Language Model
**SEM**   Standard Error of the Mean
**LFP**   Local Field Potential
**ROI**   Region of Interest
**PCA**   Principal Component Analysis
**TRF**   Temporal Response Function
**CV**   Cross Validation
**EEG**   Electroencephalography
**iEEG**   intracranial EEG
**sEEG**   stereotactic EEG
**ECoG**   Electrocorticography
**MEG**   Magnetoencephalography
**fMRI**   functional Magnetic Resonance Imaging
**STG**   Superior Temporal Gyrus
**STS**   Superior Temporal Sulcus
**ATL**   Anterior Temporal Lobe
**IPL**   Inferior Parietal Lobule
**L/IFC**   Lateral / Inferior Frontal Cortex

# Materials and Methods

## Experimental setup

*Participants and ethics.* Forty-six French-speaking individuals (22 female) undergoing evaluation of intractable epilepsy with stereotactic intracranial electrodes participated in this study. The study (NCT05217043) was approved by the National Ethics Committee and the Local IRB (Local PBN 2021 25 / National 2021-A02652-39 / IRB 21.03762.000041). Participants gave informed consent and, in the case of minors, the legal guardian also gave informed consent. Participants were informed that the study would have no impact on their clinical treatment and received no compensation. Data are located in controlled access data storage at Rothschild Foundation Hospital. All processing of data for this research was conducted exclusively by Rothschild Foundation Hospital.

*Age groups* For simplicity, participants were grouped into three age ranges (2–5, 6–11, and 12+ years) to reflect key stages in behavioral development, namely early language acquisition (2–5), onset of literacy (6–11), and the emergence of adult-level language processing (12+)[19,78].

*Inclusion criteria.* Participants were analysed only if there were no technical issues (e.g. minimum of 5 minutes of timestamped audio stimulus). One participant was incorrectly classified as a native French speaker, and thus not considered. One individual underwent two intracranial implantations, in different brain areas, six months apart and was included as two distinct participants for simplicity.

*Protocol.* Participants were invited to passively listen to an audio-book in French. Participants could listen to nine distinct segments of the audiobook whenever they liked during their stay in the Rothschild Foundation Hospital, which was on average five days. Participants were given a dedicated smartphone, which contained a unique custom-built application designed to play each segment of the audiobook and log the time of this event. The phone played audio through an external speaker. Synchronisation between brain signal and audio was performed by transmitting the audio signal augmented with triggers to the intracranial EEG (iEEG) recording system as a bipolar channel.

*Stimuli.* The audio book presented to the participants is *Le Petit Prince*, by Antoine de Saint-Exupéry[1], read in French by a female speaker[31]. There are a total of 90 minutes of audiobook divided into nine segments of ten minutes each, and containing a total of 14,353 words and 45,632 phonemes. On average, each participant listened to 40 minutes, 6,146 words, and 17,973 phonemes.

## Neural Preprocessing

*Neural recordings.* Neural recordings were acquired at 256 Hz for 32 participants, 512 Hz for 13 participants, and 2,048 Hz for one participant. We present results for LFP, which was band-pass filtered between 0.05 - 20 Hz, then downsampled to 20 Hz. To remove distributed electric fields and spatially localize the neural activity, we applied bipolar referencing within each electrode probe, based on immediate neighbors. For simplicity, we refer to these bipolar constructs as electrodes throughout the rest of the manuscript.

*Segmentation.* Except for encoding analyses, which are processed on continuous recordings, these neural signals were then segmented into epochs time-locked between -2.5 and +3.5 s relative to each phoneme or word onset. To limit the impact of potential artifacts, we clipped the neural data to five standard deviations across epochs for a given channel at each timestep. When reporting the average evoked responses, we summarize the activity by taking the absolute value of each channel, then average over all electrodes of that participant before applying baseline correction between -0.75 and -0.5 s. No baseline correction was applied for decoding.

*Electrode localization.* Electrodes were localized for each participant using the computed tomography (CT) and magnetic resonance imaging (MRI) scans taken for clinical purposes. The cortical structure was mapped by processing anatomical MRI with Freesurfer[79,80] to segment brain regions and build a 3D mesh of each participant's anatomy. The MRI and CT scans were then aligned with Curry[81] and Freesurfer, before electrodes were identified manually on the CT scans in the participant's individual coordinate space using Curry and mne[82].

*Regions of interest.* The brain areas typically recruited by speech perception have been well established in adults[83]. To summarize the effects across participants, we thus average the encoding score, or train decoding models using electrodes in ROIs. Specifically, we use the following ROIs composed of parcellations of the Desikan-Killiany[84] and Destrieux[85] Atlases provided by Freesurfer[79].

STG=[G_temp_sup-Lateral, G_temp_sup-Plan_polar, G_temp_sup-Plan_tempo, G_temp_sup-G_T_transv, G_temporal_sup, S_temporal_transverse]

STS= [S_temporal_sup]

ATL=[Pole_temporal]

IPL=[G_pariet_inf-Supramar, S_interm_prim-Jensen, G_pariet_inf-Angular]

L/IFC=[lateralorbitofrontal, frontalpole, parsopercularis, parstriangularis, parsorbitalis]

Occipital=[G_occipital_sup, G_occipital_middle, G_and_S_occipital_inf, S_calcarine, G_cuneus, S_oc_middle_and_Lunatus, S_oc_sup_and_transversal, Pole_occipital]

*Visualization.* For visualization purposes, we align the brain and the electrodes of each individual within a common MNI coordinate system. For this, we exclude the electrodes that are more than 5 mm away from the cortex, and plot the remaining ones on their nearest cortical vertex. We then morphed the individual cortical mesh to Freesurfer's *fsaverage* with the function compute_source_morph from mne[86]. We then collapse the left and right hemispheres, and overlay the electrode position onto the y and z axes. For clarity we divide the brain plots into lateral and medial views, the lateral being all electrodes less than 2 mm below the cortical surface and display only the lateral view. These visualization transformations have no impact on the analyses as they are all performed either for each electrode (encoding) or for each individual's ROI independently.

## Language models

Following a classic definition in neuroscience, we operationally define representations as linearly-readable or predictable information, on the notion that such format allows down stream neural assembly to directly use this information without additional transformation. In that view, each representations is defined by a high dimensional vectors whose bases are hereafter referred to as features.

*Phonetic features.* The onset of words and phonemes as well as the 12 phonetic features in the audio were provided in *The Little Prince* recording by Li *et al.*[31]. This set of features included 1) place of articulation information: low, mid, and high vowels and labial and velar consonants, 2) manner of articulation information: fricative, nasal, plosive, approximant, and vowel and 3) the presence of voicing and trill.

*Lexical features.* The word-level features used were Zipf frequency[34,87] and eight part of speech categories. Zipf frequency indicates, on a logarithmic scale, how common a word is in natural language. Part of speech categories were defined with spacy's `fr_core_news_lg`[88] and categorized into adverb, adjective, determinant, verb, adposition, conjunction, noun and pronouns.

*Wav2vec 2.0.* Wav2vec 2.0[35] is a self-supervised deep learning model that takes raw audio waveforms as input trained to unmask its quantized representations. This model is composed of 6 convolutional layers and 12 transformer layers. We input wav2vec 2.0 with the same audio files that were played to the participants and extracted activations of two layers, respectively position at 10% and 80% of the architecture. For the analysis of the combined effect of layers 0.8+0.1 the activations of both layers were concatenated. A Principal Component Analysis (PCA) of 50 components was taken of these embeddings, using sklearn[89] `StandardScaler` followed by `PCA`, fit per layer across the entire input data. The trained model was the "facebook/wav2vec2-base" from Huggingface[90]. The untrained wav2vec 2.0 used was the same, randomly initialized with a set seed.

*Llama 3.1* Llama 3.1 is a LLM trained on textual data. The model is composed of 32 transformer layers. We input Llama 3.1 (8B) with the text of the audio book that was played to the human participants and extracted activations from two layers positioned at 10% and 80% of its architecture. A PCA of 50 components was taken of these embeddings, fit per layer across the entire regression input data. The trained model was downloaded from "meta-llama/Llama-3.1-8B" from Huggingface[91]. The untrained Llama 3.1 used was the same, randomly initialized with a set seed.

## Encoding

Encoding analyses consist of predicting brain signals from stimulus features. Formally, let $X \in \mathbb{R}^{t \times c}$ be the brain signal recorded with $c$ electrodes at time $t$, and $Y \in \mathbb{R}^{t \times f}$ be the $f$-dimensional stimulus features at time $t$. The TRF is a linear encoding model, which consists in finding $W \in \mathbb{R}^{(\tau \cdot f) \times c}$ as follows:

$$\arg\min_W \sum_t \left\|X_t - W \cdot \bar{Y}_t\right\|^2 + \lambda_c \|W\|^2 \qquad (1)$$

where $\bar{Y} \in \mathbb{R}^{t \times (\tau \cdot f)}$ is the concatenation of $\tau$ successive time samples relative to $Y_t$, and $\lambda_c$ is an $L2$ regularization parameterized per channel.

We evaluate encoding models with a Pearson correlation $R_c$ for each channel on independent data:

$$R = \mathrm{corr}(W \cdot \bar{Y}_{\mathrm{test}}, X_{\mathrm{test}}) \qquad (2)$$

Chance level for correlation is $R = 0$, its maximum value is 1, and $R$ can only be below 0 because of noise fluctuation or if the assumptions of linear modeling (and in particular the identical distribution between the training and test sets) are violated.

Here, we fit a TRF, implemented using `mne.decoding.ReceptiveField`[32,86] and scikit-learn's *RidgeCV*, with 10 $\lambda$ value logarithmically-spaced between 0.01 and $10^7$, optimized over 20% of the training data. We use $\tau$=24 lags between -0.2 and 1 s relative to $t$.

We evaluate the TRF with Pearson correlation for each channel $R_c$ on the test set. We repeat this operation with a Cross Validation (CV) with 10 splits of each chapter of each participant using scikit-learn's *KFold* (shuffle=False). Finally, correlation scores are averaged across splits within each subject, thus leading to one score per feature and per electrode.

## Decoding

Decoding analyses consist of predicting stimulus features from epoched brain signals. Formally, we optimize a linear decoder $V_i \in \mathbb{R}^{(c \times \tau') \times f}$ across electodes at each time sample $i$ relative to stimulus onset as follows:

$$\arg\min_{V_i} \sum_t \left\|Y_t - V_i \cdot \bar{X}_{(t,i)}\right\|^2 + \lambda_{(f,i)} \|V_i\|^2 \qquad (3)$$

where $\bar{X}_{t,i}$ is the concatenated brain activity of $\tau' = 3$ lags centered around $i$. We optimize $V_i$ for each time step $i$ from -2.5 s before onset to 3.5 s after the stimulus onset.

We use scikit-learn's *RidgeCV*, with 10 $\lambda$ value logarithmically-spaced between 0.01 and $10^7$.

We then evaluate it with a Spearman correlation $R_t$ for each time sample of the test set. We repeat this operation with a 10-split CV across epochs per subject using scikit-learn's *KFold* (shuffle=False). Finally, correlation scores are averaged across splits within each subject, thus leading to one score per feature x time-sample x participant.

For each participant, we either apply decoding analyses across all electrodes, or across all electrodes of a region-of-interest.

## Statistics

*Within groups.* Except if stated otherwise, we evaluate whether correlation scores $R \in \mathbb{R}^s$ are above chance across $s$ participants, with a non-parametric one-tailed Wilcoxon test. Note that $R$ can be averaged across specific dimensions. For example, decoding scores can be evaluated across participants by averaging the $R$ scores across time samples and language features.

*Across groups.* To test whether the scores of a group of participants ($R \in \mathbb{R}^s$) is higher than those of another group ($R \in \mathbb{R}^{s'}$, with $s \neq s'$) are significantly different, we use the non-parametric one-tailed Mann-Whitney U test.

*Within participant.* To estimate the statistical effects within each subject, we use permutation analyses to compare our encoding and decoding scores to a random distribution. The random distribution is here determined by training the encoding or decoding model on all data in a chapter, then evaluating its $R$ scores on 1,000 different versions of the time-shuffled target variable $\tilde{X}_i$, e.g.:

$$R_i = \mathrm{corr}(\tilde{X}_i, W \cdot Y) \quad (4)$$

The $p$-value can then be approximated as the probability of finding random $R_i$ superior or equal to $R$.

*Correction for multiple comparisons.* To correct for multiple comparisons (e.g. across either electrodes or time samples), we apply a False Discovery Rate procedure[92], and report the corrected values. Significance is considered obtained over electrodes or time samples if the corrected $p$-value is lower than 0.05.

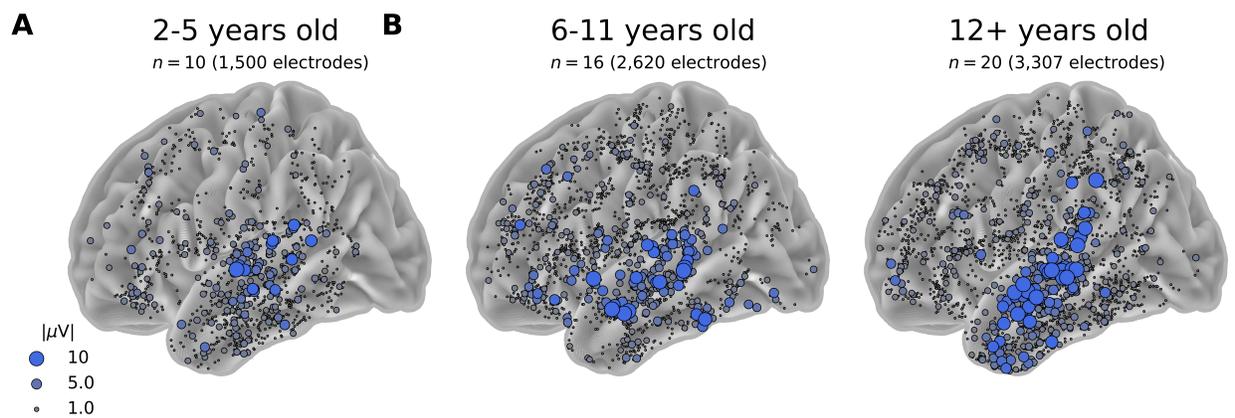

**Fig. S1| Neural responses to natural speech across development.** Average broadband response at 100 ms relative to phoneme onset.

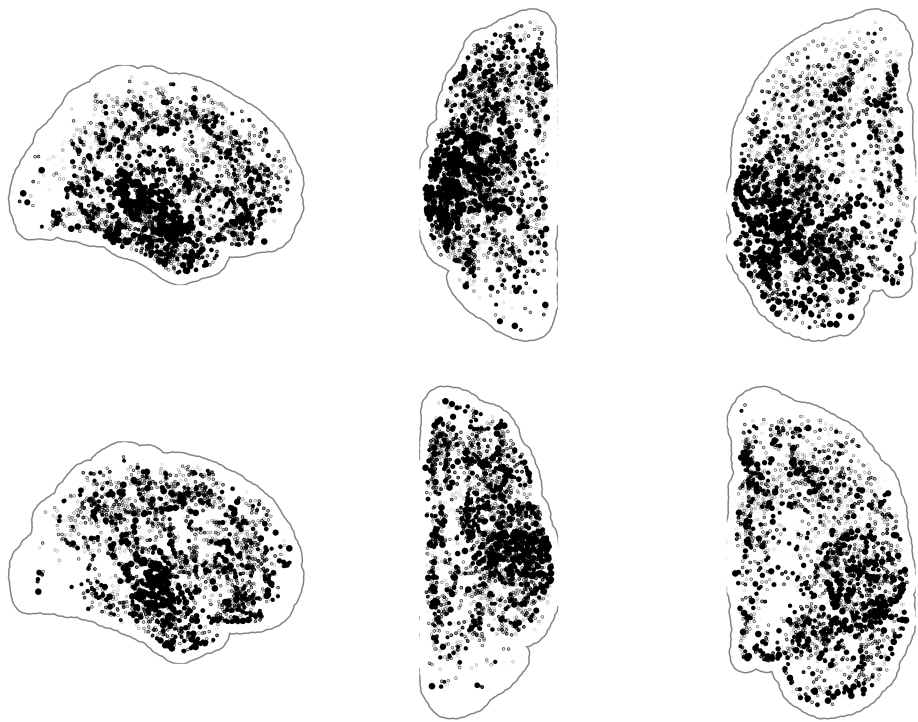

**Fig. S2| Neural responses to natural speech across development.** The broadband response per electrode at 100 ms relative to phoneme onset.

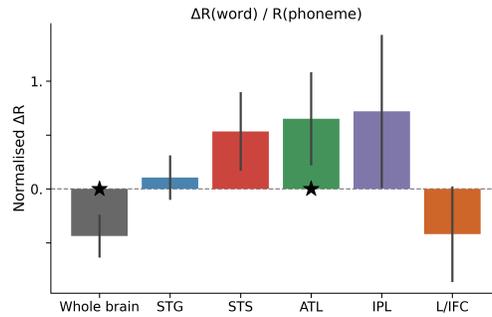

**Fig. S3| The normalised ΔR due to word level information per ROI.** The ΔR due to word level information, normalised by R(phoneme): Normalised ΔR = (R(word+phoneme) - R(phoneme)) / R(phoneme). Stars indicate ROIs with significant (p<0.05) difference from zero, using a two-sided Wilcoxon test across participants.

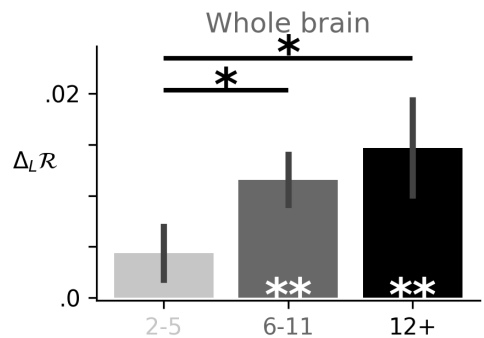

**Fig. S4| Age dependent language hierarchy modeled by wav2vec 2.0.** The age-dependent gain in encoding score offered by a later layer in wav2vec 2.0. $\Delta_L R = R_{0.8+0.1} - R_{0.1}$

16